# Model order reduction applied to heat conduction in photovoltaic modules

S.O. Ojo[1], S. Grivet-Talocia[2], M. Paggi[3,*]

[1]Department of Structural, Geotechnical and Building Engineering, Politecnico di Torino, C.so Duca degli Abruzzi 24, 10129 Torino, Italy

[2]Department of Electronics and Telecommunications, Politecnico di Torino, C.so Duca degli Abruzzi 24, 10129 Torino, Italy

[3]IMT Institute for Advanced Studies Lucca, Piazza San Francesco 19, 55100 Lucca, Italy

**ABSTRACT**

Modelling of physical systems may be a challenging task when it requires solving large sets of numerical equations. This is the case of photovoltaic (PV) systems which contain many PV modules, each module containing several silicon cells. The determination of the temperature field in the modules leads to large scale systems, which may be computationally expensive to solve. In this context, Model Order Reduction (MOR) techniques can be used to approximate the full system dynamics with a compact model, that is much faster to solve. Among the several available MOR approaches, in this work we consider the Discrete Empirical Interpolation Method (DEIM), which we apply with a suitably modified formulation that is specifically designed for handling the nonlinear terms that are present in the equations governing the thermal behaviour of PV modules. The results show that the proposed DEIM technique is able to reduce significantly the system size, by retaining a full control on the accuracy of the solution.

---

[*] Corresponding Author. E-mail: marco.paggi@imtlucca.it; Tel: +39 0583 4326 604, Fax: +39 0583 4326 565

# 1. Introduction

A photovoltaic (PV) system usually consists of an array of PV modules (e.g. 10), and each module contains several solar silicon cells (e.g. 60 or 70 in commercial modules). Each module is a layered composite such that the silicon cells are sandwiched between the different layers (see Fig. 1 for a schematic representation of a module cross-section). The computation of thermal stresses and thermo-elastic displacements is of paramount importance both during the production process, and during the operating conditions of the module. In the former case, compressive residual thermo-elastic stresses may arise and are beneficial in case of cracks within silicon due to the activation of partial crack closure [1]. In the latter, cyclic thermoelastic stresses are responsible for crack growth in silicon cells and a power-loss of the PV system in time. Moreover, thermoelastic deformation may induce failure of the busbars connecting solar cells, due to an increase in the gap between cells, as experimentally and numerically studied in [2, 3]. In all of these cases, it is important to accurately compute the temperature distribution in the plane of the solar cells [4], but also the temperature in the various layers [5] for the study of fully coupled thermomechanical problems.

So far, semi-analytical and numerical solutions [2, 3] for the assessment of the change in the gap between solar cells have been proposed by assuming a uniform temperature field across the module, which is an assumption holding for stationary conditions. In reality, temperature contour plots obtained from finite element thermal analysis [6] show that there is a temperature gradient across each layer, with the regions near the frame being significantly cooler, while the temperature distribution across the cells in the centre of the module is found to be quite uniform. In addition, existence of cracks in the silicon cells may induce a localized temperature increment (hot spots) in the region near the cracks due to a localized electric resistance [7]. Moreover, transient regimes, such as those taking place in accelerated environmental tests within climate chambers, or under operating conditions, have only marginally been investigated due to the inherent complexity related to the very different thicknesses of the layers composing a PV module. In such cases, accurate predictions require the solution of large systems of equations resulting from the finite element or finite difference approximation of the (nonlinear) partial differential equation governing the problem of heat conduction. Suitable techniques for reducing the computational requirements for such simulations are therefore highly desirable.

The purpose of this study is to formulate a thermal model of a PV system, and to identify an appropriate class of Model Order Reduction (MOR) techniques, capable of approximating the full system by a compact dynamical model characterized by a much smaller number of degrees of freedom (state variables), and whose numerical solution requires a significantly reduced time with respect to the full system.

Choosing a particular MOR technique that is suitable for a given problem depends on several factors, such as type of system (linear or nonlinear), number and structure of equations to be solved for and, in case of nonlinear systems, the degree of nonlinearity. Model order reduction is a mature field when applied to linear systems, and several excellent books on the subject are available [8-11]. Many techniques are available for linear order reduction, including Krylov subspace projection based on orthogonal Arnoldi [12, 13] or biorthogonal Lanczos [14] processes, principal components analysis and balanced truncation [15], Hankel norm approximation [16], and singular value decomposition (SVD) based methods, which include Proper Orthogonal Decomposition (POD) in its many variants [17]. Many extensions to nonlinear system are also available, see e.g. [18-22], which combine system projection or truncation with suitable approximations of the nonlinear terms.

In this work, we propose a POD projection based approach for the reduction of the model dynamic order, combined with a Discrete Empirical Interpolation Method (DEIM) [22] for handling the nonlinear terms. We show that this approach is ideally suited to the inherent structure of the PV discretized heat transfer equations obtained through a finite difference scheme, allowing for massive order reduction at almost no loss of accuracy. A systematic validation study is conducted on a PV template system, showing the influence of the various parameters governing the reduction scheme. Numerical results confirm the excellent scalability of the proposed technique.

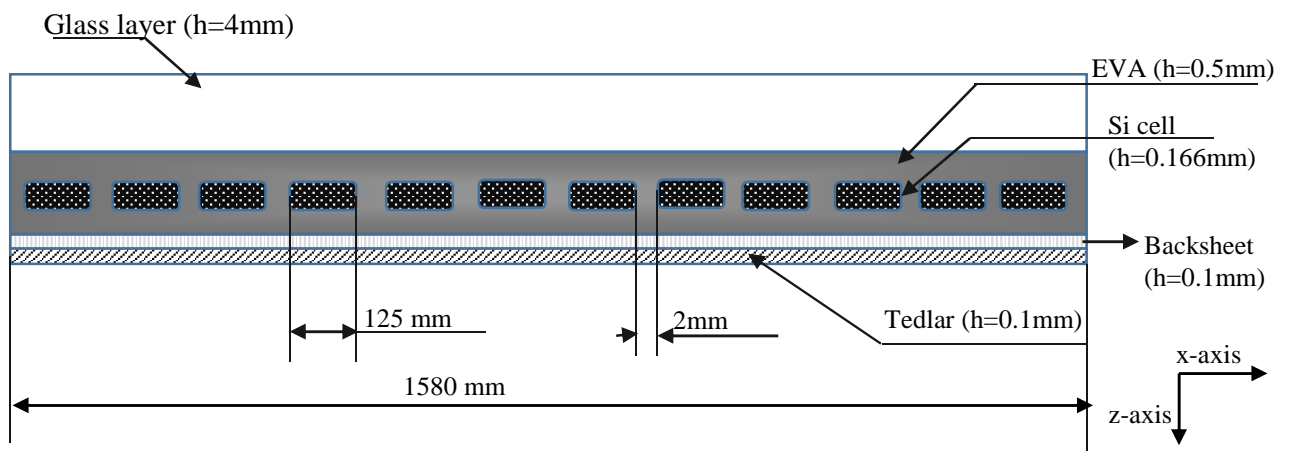

Figure 1. A sketch of a cross-section of a PV module, not in scale. For the actual value of the thicknesses, see the labels in the figure.

## 2. Formulation of the differential problem and finite difference approximation

A 2D thermal model of a PV module is proposed here based on the work by Jones [23]. We consider a PV module containing 12 silicon cells embedded in a composite made of glass, EVA, Silicon, EVA, backsheet and tedlar layers with the properties described in Tab. 1 [24]. Although solar cells are separated from each other by a small amount of EVA in their plane, in this work we slightly simplify the structure and we consider all layers as uniform in the x and y directions, see Fig. 1. Further, we consider the y direction as infinite.

Under the above assumptions, the following general 2D heat equation for the composite PV panel holds:

$$C\frac{\partial T}{\partial t} = \lambda_x \frac{\partial^2 T}{\partial x^2} + \lambda_z \frac{\partial^2 T}{\partial z^2} + H - G \qquad (1)$$

where $T(x,z,t)$ represents the space and time dependent temperature profile of the module. $C(x,z)$ is an equivalent volumetric heat capacity (J/m³K), which is equal to an equivalent mass density times the equivalent specific heat capacity (C =$\rho \cdot c_p$) taking into account the composite structure of the laminate, see next section. The function $H(x,z,t)$ represents the heat losses by radiation and convection taking place at any point within the PV module, and $G(x,z,t)$ is the electrical energy generated by the cell layer. The coefficients $\lambda_x(x,z)$ and $\lambda_z(x,z)$ are the thermal conductivities in the x and z directions respectively. According to Fourier's law of heat conduction, the heat flows in the x and z direction are related to these thermal conductivities by

$$q_z = -\lambda_z \frac{\partial T}{\partial z}, \qquad q_x = -\lambda_x \frac{\partial T}{\partial x} \qquad (2)$$

Substituting (2) into (1), we have:

$$C\frac{\partial T}{\partial t} = -\frac{\partial q_x}{\partial x} - \frac{\partial q_z}{\partial z} + H - G \qquad (3)$$

Using now a finite difference (FD) discretization scheme defined by grid spacing $\Delta x_i$ and $\Delta z_j$ in the x and z-direction, respectively, with associated discretization indices $i$ for $1 \leq i \leq l$ and $j$ for $1 \leq j \leq s$ (see Fig. 2), we can rephrase (3) as

$$C_{i,j}\frac{dT_{i,j}}{dt} = \frac{q_{i-\frac{1}{2},j} - q_{i+\frac{1}{2},j}}{\Delta x_i} + \frac{q_{i,j-\frac{1}{2}} - q_{i,j+\frac{1}{2}}}{\Delta z_j} + H_{i,j} - G_{i,j} \qquad (4)$$

| Layer | Thickness (mm) | Thermal conductivity (W/m°K) | Density (Kg/m³) | Specific heat capacity (J/Kg°K) |
|---|---|---|---|---|
| Glass | 4 | 1.8 | 3000 | 500 |
| EVA | 0.5 | 0.35 | 960 | 2090 |
| PV Cells | 0.166 | 148 | 2330 | 677 |
| EVA | 0.5 | 0.35 | 960 | 2090 |
| Back contact | 0.1 | 237 | 2700 | 900 |
| Tedlar | 0.1 | 0.2 | 1200 | 1250 |

Table 1. Material properties of the layers of the PV module

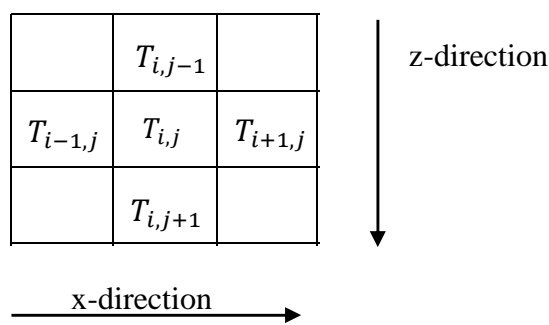

Figure 2. Finite difference discretization of the PV module in its plane.

where $q_{i-\frac{1}{2},j}$ and $q_{i+\frac{1}{2},j}$ are the heat flows through the left and right boundaries of element $(i,j)$, and $q_{i,j-\frac{1}{2}}$ and $q_{i,j+\frac{1}{2}}$ are the heat flows through the upper and lower boundaries of the element in its plane. Multiplying (4) by the area $A_{i,j} = \Delta x_i \Delta z_j$ of each grid element in the FD discretization leads to

$$C_{i,j} A_{i,j} \frac{\partial T_{i,j}}{\partial t} = Q_{i-\frac{1}{2},j} - Q_{i+\frac{1}{2},j} + Q_{i,j-\frac{1}{2}} - Q_{i,j+\frac{1}{2}} + A_{i,j} H_{i,j} - A_{i,j} G_{i,j}$$
(5)

where $Q$ represents, consistently with energy conservation principles, the heat flow between adjacent cells, which can be further expressed as $Q = K \Delta T$ [24], where $\Delta T$ is the temperature change between the two cells, and $K$ is the corresponding thermal conductance. The latter is a function of the equivalent thermal conductivities of the two cells and the width and length of the elements, i.e. $\Delta z_j$ and $\Delta x_i$.

## 2.1 Thermal conductances and heat flows

The discretized thermal conductances $K_{i,j}$ (W/mK) provide information on the thermal coupling between the elements in the discretization of the PV module. Assuming perfect bonding at the various interfaces between the layers, the thermal conductance per unit length in the x-direction between cells $(i-1,j)$ and $(i,j)$ is given by [25]

$$K_{i-\frac{1}{2},j} = \frac{\Delta z_j}{\Delta x_{i-1}/2\lambda_{x\,i-1,j} + \Delta x_i/2\lambda_{x\,i,j} + R_{i-\frac{1}{2},j}}$$
(6)

where R is the thermal resistance at the interface between the elements. Since in the present approximation, the PV layers are uniform in the x-direction, the thermal conductivity $\lambda_x$ does not change in the x direction and we have $R_{i-\frac{1}{2},j} = 0$ and we can simplify notation as $\lambda_{x\,i-1,j} = \lambda_{x\,i,j} = \lambda_j$. This assumption is still reasonable for the silicon cell layer, since the cells are separated from each other by a small amount of EVA (2mm), much smaller than the lateral size of each silicon cell (125mm). Grid spacing $\Delta x$ is also considered to be constant in the x direction. Therefore (6) becomes

$$K_{i-\frac{1}{2},j} = \frac{\Delta z_j}{\Delta x/2\lambda_j + \Delta x/2\lambda_j} = \lambda_j \frac{\Delta z_j}{\Delta x}$$
(7)

and $K_{i-\frac{1}{2},j} = K_{i+\frac{1}{2},j} = K_j$ due to material homogeneity in the x-direction.

In the z-direction, we have:

$$K_{j-\frac{1}{2}} = \frac{\Delta x}{\Delta z_{j-1}/2\lambda_{z\,j-1} + \Delta z_j/2\lambda_{z\,j} + R_{j-\frac{1}{2}}}$$
(8)

$$K_{j+\frac{1}{2}} = \frac{\Delta x}{\Delta z_{j+1}/2\lambda_{z\,j+1} + \Delta z_j/2\lambda_{z\,j} + R_{j+\frac{1}{2}}}$$

At the top and bottom boundary elements of the PV module, $(i,1)$ and $(i,s)$, we have

$$K_{\frac{1}{2}} = \frac{\Delta x}{\Delta z_1/2\lambda_{z\,1} + R_{\frac{1}{2}}}$$
(9)

$$K_{s-\frac{1}{2}} = \frac{\Delta x}{\Delta z_s/2\lambda_{z\,s} + R_{s-\frac{1}{2}}} \qquad (10)$$

where $R_{\frac{1}{2}}$ and $R_{s-\frac{1}{2}}$ are the thermal resistances between the top and bottom elements and the free surfaces.

From (5), the heat flows through the left and the right boundaries of the element $(i,j)$ are thus defined as

$$Q_{i-\frac{1}{2},j} = K_j(T_{i-1,j} - T_{i,j}) \qquad (11)$$

$$Q_{i+\frac{1}{2},j} = K_j(T_{i,j} - T_{i+1,j}) \qquad (12)$$

whereas the heat flows through the lower and upper boundary of the element $(i,j)$ are

$$Q_{i,j-\frac{1}{2}} = K_{j-\frac{1}{2}}(T_{i,j-1} - T_{i,j}) \qquad (13)$$

$$Q_{i,j+\frac{1}{2}} = K_{j+\frac{1}{2}}(T_{i,j} - T_{i,j+1}) \qquad (14)$$

## 2.2 Boundary conditions

In this study, a constant (Dirichlet) temperature is applied to the right and left boundary of the module. Thus, the heat flow at the left and right boundary elements of the PV module $(1,j)$ and $(l,j)$ are

$$Q_{\frac{1}{2},j} = K_j(T_{b1} - T_{1,j}) \qquad (15)$$

$$Q_{l-\frac{1}{2},j} = K_j(T_{l,j} - T_{b2}) \qquad (16)$$

Where $T_{b1}$ and $T_{b2}$ are the fixed temperatures imposed at the right and left of the module. In all subsequent simulations, we will set them equal to 343 K and 313 K in order to simulate a distinct differential temperature profile from one end of the PV module to the other.

The heat flow at the top and bottom boundary elements of the PV module $(i,1)$ and $(i,s)$ are instead

$$Q_{i,\frac{1}{2}} = K_{\frac{1}{2}}(T_{sky} - T_{i,1}) \qquad (17)$$

$$Q_{i,s-\frac{1}{2}} = K_{s-\frac{1}{2}}(T_{i,s} - T_{sky}) \qquad (18)$$

where $T_{sky}$ is the temperature of the sky.

## 2.3 Heat loss

The heat loss, which varies through the layer thickness of the module, is given by the sum of the following contributions [23]

$$H(x,z,t) = q_{lw}(x,z,t) + q_{sw}(x,z,t) + q_{conv}(x,z,t) \qquad (19)$$

where the long wave, short wave and convection heat transfers are denoted by $q_{lw}$, $q_{sw}$ and $q_{conv}$ respectively.

The short wave radiation heat transfer of a body of area A is given by

$$q_{sw} = A\alpha\Phi \qquad (20)$$

where $\alpha$ and $\Phi$ are the absorptivity of the material and the total incident irradiance input to the module surface, respectively. The long wave radiation heat transfer is given by the Stefan Boltzmann law

$$q_{lw} = \sigma \varepsilon T^4 \tag{21}$$

where $\sigma$ is the Boltzmann's constant ($5.607 \times 10^{-8}$ Js$^{-1}$m$^{-2}$K$^{-4}$) and $\varepsilon$ is the emissivity of the body. It is assumed that the net long wave exchange is negligible for the rear of the module. Thus, it is only necessary to calculate the long wave exchange from the surface of the module. The net long wave radiation exchange between two surfaces x and y is given by [23]

$$q_{lw_{xy}} = A_x F_{xy}(L_x - L_y) = A_y F_{xy}(L_y - L_x) \tag{22}$$

Here $L_x$ and $L_y$ are long wave irradiance per unit area for surface x and y respectively which are given by

$$L_x = \sigma \varepsilon_x T_x^4 \text{ and } L_y = \sigma \varepsilon_y T_y^4$$

where $F_{xy}$ is the view factor, a fraction of the radiation leaving surface x that reaches surface y.

A tilted module surface not overlooked by adjacent buildings at an angle $\theta$ from the horizontal has a view factor of $\frac{(1+\cos\theta)}{2}$ for the sky and $\frac{(1-\cos\theta)}{2}$ for the horizontal ground [23]. Thus, inserting the view factor coefficient for sky and ground into $L_x$ we have

$$L_x = \sigma \frac{(1+\cos\theta)}{2} \varepsilon_{sky} T_{sky}^4 + \sigma \frac{(1-\cos\theta)}{2} \varepsilon_{ground} T_{ground}^4 \tag{23}$$

$$L_y = \sigma \varepsilon_{mod} T^4 \tag{24}$$

where $\varepsilon_{mod}$ is the module emissivity.

Substituting (23) and (24) into (22), we have

$$q_{lw} = A\sigma \left( \frac{(1+\cos\theta)}{2} \varepsilon_{sky} T_{sky}^4 + \frac{(1-\cos\theta)}{2} \varepsilon_{ground} T_{ground}^4 - \varepsilon_{mod} T^4 \right) \tag{25}$$

Further, $T_{sky} = T_{ambient} - \delta T$ for clear sky condition in which $\delta T = 20K$ and $T_{sky} = T_{ambient}$ for overcast condition.

The convection heat transfer is related to the temperature gap between the upper part of the solar panel and the ambient [23]

$$q_{conv} = -A(h_{c,forced} + h_{c,free})(T - T_{ambient}) \tag{26}$$

where $h_{c,forced}$ and $h_{c,free}$ in W/m² K are the forced and free convection heat transfer coefficients which depend on the wind speed.

Collecting all the heat loss contributions together, we finally obtain

$$H_{i,j} = A_{i,j}[\alpha_{i,j}\Phi + \sigma\left(\frac{(1+\cos\theta)}{2}\varepsilon_{sky}T_{sky}^4 + \frac{(1-\cos\theta)}{2}\varepsilon_{ground}T_{ground}^4 - \varepsilon_{i,j}T_{i,j}^4\right) - (h_{c,forced} + h_{c,free})(T_{i,j} - T_{ambient})]$$ (27)

where $\alpha_{i,j}$ and $\varepsilon_{i,j}$ denote the absorptivity and emissivity coefficients of the discretized elements in the different layers.

*2.4 Power generated by the PV Cell*

The power generated by the PV cell at location $(i,j)$ can be estimated as [23]

$$G_{i,j} = C_{FF}\frac{E\,ln(\gamma E)}{T_{i,j}}$$ (28)

where $C_{FF}$ is the fill factor model constant (1.22 K m²) and $E(t)$ in $W/m^2$ is the incident irradiance input through the thickness of the PV module. The constant $\gamma$ is equal to $10^6$ m²/W. It should be noted that the power generated by the discretized PV cells in (28) is non-zero only for the silicon cell layer.

The incident irradiance input into the system is obtained from experimental data [23]. To obtain a validated result of the reduced order model to be derived in Sec. 3, a minute by minute irradiance input obtained from the solar resource and meteorological assessment project website (http://www.nrel.gov/midc/kalaeloa_oahu/) will be used. The plots for irradiance for a period of 30 min are shown below in Fig. 3.

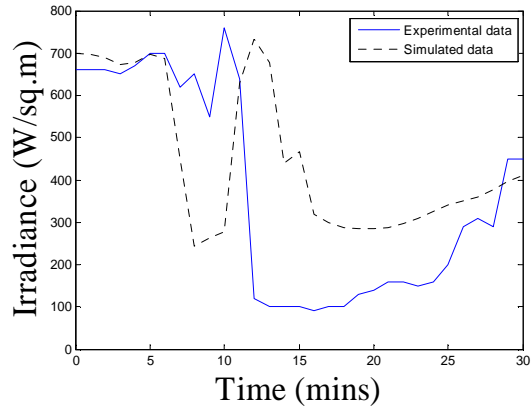

Figure 3. Experimental and simulated irradiance input (from 09:52-10:22, 11/01/2011)

*2.5 System of nonlinear ODEs for the PV module*
Considering all the relations established in Sec. 2, the discretized thermal equation (5) can be rewritten after substituting the corresponding expressions for Q, $H$ and $G$ as

$$C_{i,j}A_{i,j}\frac{\partial T_{i,j}}{\partial t} - K_j T_{i+1,j} - K_j T_{i-1,j} + T_{i,j}\left(2K_j + K_{i,j+\frac{1}{2}} + K_{i,j-\frac{1}{2}}\right) - K_{j+\frac{1}{2}}T_{i,j+1} - K_{j-\frac{1}{2}}T_{i,j-1} - A_{i,j}[\alpha_{i,j}\Phi + \sigma\left(\frac{(1+\cos\theta)}{2}\varepsilon_{sky}T_{sky}^4 + \frac{(1-\cos\theta)}{2}\varepsilon_{ground}T_{ground}^4 - \varepsilon_{i,j}T_{i,j}^4\right) - (h_{c,forced} + h_{c,free})(T_{i,j} - T_{ambient})] = -A_{i,j}C_{FF}\frac{E\,ln(\gamma E)}{T_{i,j}}$$

(29)

At the left and right boundaries of the PV module, $Q_{i-\frac{1}{2},j}$ and $Q_{i+\frac{1}{2},j}$ from (5) are replaced by $Q_{\frac{1}{2},j}$ and $Q_{l-\frac{1}{2},j}$ respectively from (15) and (16), while at the top and bottom boundaries of the PV module, $Q_{i,j-\frac{1}{2}}$ and $Q_{i,j+\frac{1}{2}}$ are replaced by $Q_{i,\frac{1}{2}}$ and $Q_{i,s-\frac{1}{2}}$ respectively from (17) and (18).

The discretized thermal equation (29) can finally be written in a compact matrix form

$$CA\frac{dT(t)}{dt} = \bar{K}T(t) + AI(T(t), E(t)) \tag{30}$$

or, in explicit form, as

$$\frac{dT(t)}{dt} = \underbrace{KT(t)}_{\text{linear term}} + \underbrace{F(T(t), E(t))}_{\text{Non-linear term}} \tag{31}$$

where $K = (CA)^{-1} \cdot (\bar{K})$ and $F = C^{-1}I$. The independent variable $t \in [0, h]$ denotes time, and $T(t) = [T_1(t), \ldots, T_n(t)]^T \in \mathbb{R}^n$ is the unknown temperature vector for all the elements in the FD discretization, where we use a single subscript with $n$ denoting the total number of nodes. $E(t)$ is the time-varying irradiance input to the system, the matrix $K \in \mathbb{R}^{n \times n}$ contains constants and $F(T(t), E(t))$ is a nonlinear function evaluated at $T(t)$ component-wise i.e.

$$F = [F_1, \ldots, F_n]^T \tag{32}$$

### 2.6 Reference solution for the complete system

Before applying our proposed model order reduction approach, we derive a reference solution for (31) by direct time discretization. A backward finite difference scheme is selected to solve the thermal problem to avoid any convergence issues associated with explicit methods in terms of choice of time step. The numerical method is implemented in Matlab. A uniform discretization of the module in the x-direction is adopted with $l$=361 grid points, while there are $s$=6 strips in the z-direction with different thicknesses so that $n = s \times l = 2166$. The solution of this problem is performed for $n_s$=186 time steps, each step representing 10 s of physical time. Figure 4(a) shows the temperature profile for node 741 in the silicon layer for all the 186 time steps. The temperature along the silicon layer vs. position at the last time step is shown in Fig. 4(b). As it can be seen, the transient regime is quite evident and the temperature in the silicon cell layers is significantly different from cell to cell.

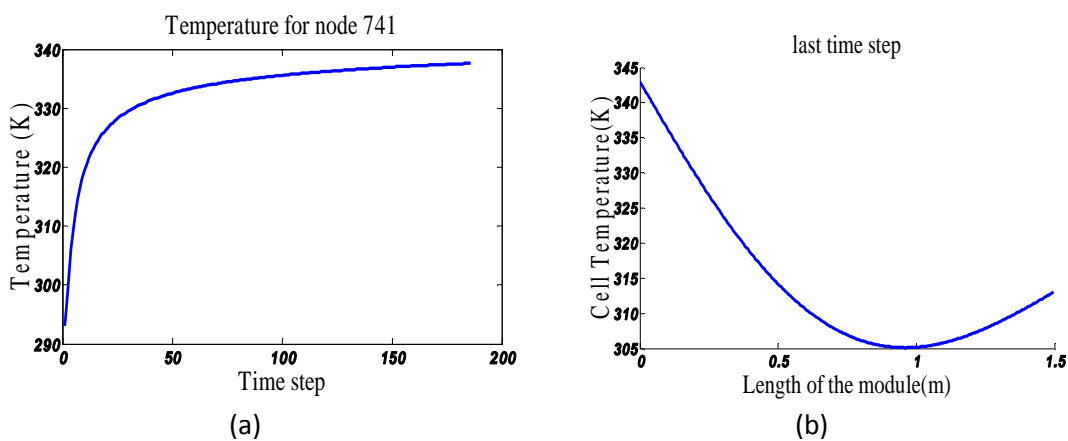

(a) (b)

Figure 4. Temperature profile of the full system for (a) a node within the silicon layer vs. time step and (b) along the silicon layer at the last time step of the simulation.

## 3. Model order reduction

The direct numerical simulation of (31) may be quite demanding in terms of computing resources, especially in view of its extension to a full 3D geometry. For this reason, we investigate in this work a MOR technique, with the objective of approximating the large-scale system (31) with a lower order compact dynamical model, that is able however to preserve accuracy in its input/output transient response. The two key aspects of proposed MOR approach are: i) a massive reduction in the degrees of freedom (states), and ii) an accurate representation of the nonlinear terms that influence the heat exchange of the structure. These two aspects are analysed in detail.

*3.1 System projection*

The reduction in the degrees of freedom is here performed through a standard projection approach. The vector $\boldsymbol{T}$ collecting all $n$ cell temperatures is approximated as a linear superposition of a small number $k$ of "basis vectors", which span a reduced order subspace. More precisely, we consider the representation $\boldsymbol{T} \approx \boldsymbol{V}_k \widetilde{\boldsymbol{T}}$, where $\widetilde{\boldsymbol{T}} \in \mathbb{R}^k$ is a reduced temperature vector collecting the coefficients of $\boldsymbol{T}$ into a reduced basis, defined by the columns of matrix $\boldsymbol{V}_k \in \mathbb{R}^{n \times k}$. We consider an orthonormal basis, so that $\boldsymbol{V}_k^T \boldsymbol{V}_k = \boldsymbol{I} \in \mathbb{R}^{k \times k}$ ($k \ll n$) with $\boldsymbol{I}$ an identity matrix. Introducing the above reduced expression for $\boldsymbol{T}$ into (31), we have

$$\boldsymbol{V}_k \frac{d\widetilde{\boldsymbol{T}}(t)}{dt} \approx \boldsymbol{K}\boldsymbol{V}_k \widetilde{\boldsymbol{T}}(t) + \boldsymbol{F}(\boldsymbol{V}_k \widetilde{\boldsymbol{T}}(t), E(t)) \tag{33}$$

Projecting now these equations along the subspace generated by $\boldsymbol{V}_k$ leads to

$$\frac{d\widetilde{\boldsymbol{T}}(t)}{dt} \approx \underbrace{\boldsymbol{V}_k^T \boldsymbol{K}\boldsymbol{V}_k}\, \widetilde{\boldsymbol{T}}(t) + \boldsymbol{V}_k^T \boldsymbol{F}(\boldsymbol{V}_k \widetilde{\boldsymbol{T}}(t), E(t))$$

with $\boldsymbol{V}_k^T \boldsymbol{K}\boldsymbol{V}_k = \widetilde{\boldsymbol{K}}$ and where $\widetilde{\boldsymbol{K}} \in \mathbb{R}^{k \times k}$

The reduced form of the thermal equation (31) reads

$$\frac{d\widetilde{\boldsymbol{T}}(t)}{dt} = \widetilde{\boldsymbol{K}}\widetilde{\boldsymbol{T}}(t) + \boldsymbol{V}_k^T \boldsymbol{F}\big(\boldsymbol{V}_k \widetilde{\boldsymbol{T}}(t),\ E(t)\big) \tag{34}$$

The above system represents a reduced order model, since its main variables are the coefficients of a reduced basis. In order to determine $\boldsymbol{V}_k$, we use a Proper Orthogonal Decomposition (POD), which extracts the basis vectors from the actual transient solution of the full system by means of a truncated singular value decomposition. In particular, we collect the $n_s$ snapshots $\boldsymbol{T}(t_h)$ obtained from the full solution of the system at discrete time steps of size $h$ in the following snapshot matrix

$$\boldsymbol{S} = [\boldsymbol{T}(t_1), \ldots, \boldsymbol{T}(t_{ns})] \tag{35}$$

and we apply the POD algorithm below

**INPUT:** $S = [T(t_1), \ldots, T(t_{ns})] \subset \mathbb{R}^{n \times ns}$
**OUTPUT:** $V_k = [v_1, \ldots, v_k] \in \mathbb{R}^{n \times k}$

1. Form the shapshot matrix $S = \{T(t_1), \ldots, T(t_{ns})\}$
2. Perform the singular value decomposition $T = V\Sigma W^T$ to produce orthogonal matrices $V = [v_1, \ldots, v_r] \in \mathbb{R}^{n \times r}$ and $W = [w_1, \ldots w_r] \in \mathbb{R}^{n_s \times r}$ and diagonal matrix $\Sigma = \text{diag}(\sigma_1, \ldots, \sigma_r) \in \mathbb{R}^{r \times r}$ where r is the rank of $S$.
3. Set a threshold to pick the $k$ highest modes from the diagonal matrix $\Sigma$
4. Pick the columns in matrix $V$ which correspond to the modes selected in 3 to generate the POD basis $V_k = [v_1, \ldots, v_k] \in \mathbb{R}^{n \times k}$

Note that in the present case we choose $n_s = 186$, as the total number of time steps in the full solution. The choice of $n_s$ should be carefully considered since it can strongly influence the accuracy of the approximation and the computational cost, as shown later.

*3.2 Discrete Empirical Interpolation Method (DEIM)*

System (34) is a reduced order model, but the evaluation of the nonlinear term still requires the mapping $V_k$ to the full-size space. The DEIM approach is used here to further approximate the nonlinear terms, thus reducing the computational cost associated with the simulation of the reduced model. According to [10], we write the nonlinear term of (34) as

$$N(\widetilde{T}) = \underbrace{V_k^T}_{k \times n} \underbrace{F(V_k\widetilde{T}(t), E(t))}_{n \times 1}$$

(36)

and we define

$$f(t) = F(V_k\widetilde{T}(t), E(t)) \tag{37}$$

The basic idea is to approximate $f(t)$ by projecting it onto the subspace spanned by a suitable set of $m \ll n$ basis vectors $\mathbf{u}_1, \ldots, \mathbf{u}_m$ via
$$f(t) \approx \mathbf{U}\mathbf{c}(t) \tag{38}$$
where $\mathbf{U} = [\mathbf{u}_1, \ldots, \mathbf{u}_m] \in \mathbb{R}^{n \times m}$. The corresponding coefficient vector $\mathbf{c}(t)$ is determined by selecting $m$ significant rows from the overdetermined system (38). This can be achieved by considering the mapping matrix
$$\mathbf{P} = [\mathbf{e}_{\varrho_1}, \ldots, \mathbf{e}_{\varrho_m}] \in \mathbb{R}^{n \times m} \tag{39}$$

Where $\mathbf{e}_{\varrho_i} = [0, \ldots, 0, 1, 0, \ldots, 0]^T \in \mathbb{R}^n$ is the $\varrho_i$th column of the identity matrix $\mathbf{I} \in \mathbb{R}^{n \times n}$ for $i = 1, \ldots, m$. The coefficient $\mathbf{c}(t)$ can thus be determined by inverting system

$$\mathbf{P}^T f(t) = (\mathbf{P}^T \mathbf{U})\mathbf{c}(t) \tag{40}$$

Provided $\mathbf{P}^T\mathbf{U}$ is non-singular, the final approximation of (38) is
$$f(t) \approx \mathbf{U}\mathbf{c}(t) = \mathbf{U}(\mathbf{P}^T\mathbf{U})^{-1}\mathbf{P}^T f(t) \tag{41}$$
Since $f(t) = F(V_k \widetilde{T}(t), E(t))$, (41) can thus be written as
$$F(V_k \widetilde{T}(t), E(t)) \approx \mathbf{U}(\mathbf{P}^T\mathbf{U})^{-1}\mathbf{P}^T F(V_k \widetilde{T}(t), E(t)) \tag{42}$$
Eq. (42) ensures that the nonlinear function $F$ is evaluated for the full system and then interpolated by matrix $\mathbf{P}$, an operation which still shows the dependence of the reduced system on the complete system size. To avoid this dependence, DEIM interpolates the input vector of the nonlinear function $F$ and then evaluates $F$ component-wise at its interpolated input vector. Based on this, (42) can be written as
$$F(V_k \widetilde{T}(t) E(t)) \approx \mathbf{U}(\mathbf{P}^T\mathbf{U})^{-1} \widetilde{F}(\mathbf{P}^T V_k \widetilde{T}(t), E(t)) \tag{43}$$
Where $\widetilde{F}$ denotes the selected components of $F$. This approximation is particularly effective when the full nonlinear function $F$ is evaluated independently for each component of its vector argument, as in present FD formulation. The nonlinear term in (36) can now be represented as
$$N(\widetilde{T}(t)) = \underbrace{V_k^T \mathbf{U}(\mathbf{P}^T\mathbf{U})^{-1}}_{k \times m} \underbrace{\widetilde{F}(\mathbf{P}^T V_k \widetilde{T}(t), E(t))}_{m \times 1} \tag{44}$$
Now, to evaluate $N(\widetilde{T})$ in (44), we must specify the projection basis $[\mathbf{u}_1, \ldots, \mathbf{u}_m]$ and the interpolation indices $[\varrho_1, \ldots, \varrho_m]$. We can obtain the basis $[\mathbf{u}_1, \ldots, \mathbf{u}_m]$ by applying the above described POD scheme to the matrix collecting the nonlinear snapshots $F = \{F(T(t_1)), \ldots, F(T(t_{ns}))\}$ resulting from a direct evaluation of the nonlinear function of the full system at different time steps, and then using the DEIM algorithm described in [22]. The following implementation is used to iteratively construct the basis vectors and the set of interpolation indices.

**INPUT:** $\{u_i\}_{i=1}^m \subset \mathbb{R}^n$ linearly independent
**OUTPUT:** $\vec{\varrho} = [\varrho_1, \ldots, \varrho_m] \in \mathbb{R}^m$
1. $[|\rho|, \varrho_1] = \max\{|\mathbf{u}_1|\}$
2. $\mathbf{U} = |\mathbf{u}_1|, \mathbf{P} = [\mathbf{e}_{\varrho_1}], \vec{\varrho} = [\varrho_1]$
3. for $i = 2$ to m do
4. Solve $(\mathbf{P}^T\mathbf{U})\mathbf{c} = \mathbf{P}^T u_i$ for $\mathbf{c}$
5. $\mathbf{r} = u_i - \mathbf{U}\mathbf{c}$
6. $[|\rho|, \varrho_i] = \max\{|\mathbf{r}|\}$
7. $\mathbf{U} \leftarrow [\mathbf{U}\; u_i], \mathbf{P} = [\mathbf{P}\; \mathbf{e}_{\varrho_i}], \vec{\varrho} = \begin{bmatrix}\vec{\varrho}\\ \varrho_i\end{bmatrix}$

**end for**

*3.3 Modification of DEIM formulation*
To control the accuracy of the reduced system more efficiently, we notice that: i) there are two nonlinear terms with different characteristics in the thermal formulation of the PV module, and that ii) these two terms influence different layers of the PV module. In fact, since we assumed that the net long wave exchange for the rear of the module is negligible (see section 2.3), the heat loss term in the thermal system formulation has most impact on the surface of the PV module. On the other hand, the power output is generated only by the silicon cell (third layer). On this note, the DEIM operation is here performed separately for the two nonlinear terms using two different sets of snapshots. Accordingly, we have Eq. (43) respectively expressed for the two nonlinear terms in the reduced system as

$$F_1(V_k \widetilde{T}(t), E(t)) \approx \mathbf{U}_1(\mathbf{P}_1^\mathrm{T}\mathbf{U}_1)^{-1}\widetilde{F}_1(\mathbf{P}_1^\mathrm{T} V_k \widetilde{T}(t), E(t)) \qquad (45)$$

$$F_2(V_k \widetilde{T}(t), E(t)) \approx \mathbf{U}_2(\mathbf{P}_2^\mathrm{T}\mathbf{U}_2)^{-1}\widetilde{F}_2(\mathbf{P}_2^\mathrm{T} V_k \widetilde{T}(t), E(t)) \qquad (46)$$

Finally, the nonlinear term in (36) can now be represented as

$$N(\widetilde{T}) = \underbrace{V_k^T \mathbf{U}_1(\mathbf{P}_1^\mathrm{T}\mathbf{U}_1)^{-1}}_{k \times m_1} \underbrace{\widetilde{F}_1(\mathbf{P}_1^\mathrm{T} V_k \widetilde{T}(t) E(t))}_{m_1 \times 1} + \underbrace{V_k^T \mathbf{U}_2(\mathbf{P}_2^\mathrm{T}\mathbf{U}_2)^{-1}}_{k \times m_2} \underbrace{\widetilde{F}_2(\mathbf{P}_2^\mathrm{T} V_k \widetilde{T}(t), E(t))}_{m_2 \times 1}$$

$$(47)$$

With this modification, the interpolation of the nonlinear terms can be handled independently, enabling a finer control on reduced system complexity and efficiency.

## 4. Numerical results

In this section, we assess the accuracy of the proposed reduced modeling scheme by comparing the responses of the compact model and the original system. In particular, we investigate the convergence of the reduced system as a function of the three parameters that measure its complexity, namely the size $k$ of the reduced basis used in the state-space (linear) projection, and the two orders $m_1$, $m_2$ of the nonlinear interpolation.

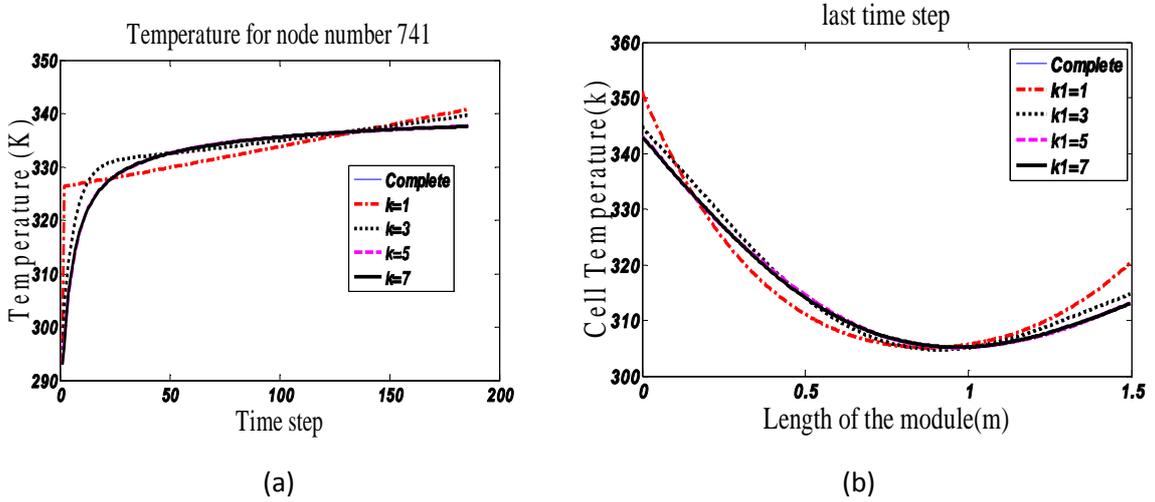

(a) (b)

Figure 5. Convergence of reduced solution to the full solution with increasing order of $k$ and fixed interpolation points $m_1=m_2=$ 3 for (a) a specific point in the silicon layer, and (b) for the entire silicon layer.

As shown in Fig. 5, the cell temperatures at the last time step of the iterative solution of the full system and of the reduced system are in fair good agreement by increasing the order $k$ of the reduced system. As the order of the reduced system increases from k=1 to k=7, the approximation of the reduced system approaches the exact value of the complete system. At k=7, the reduced system approximation fits well the complete system such that further increasing the order of the reduced system does not change the result significantly.

It is noticed that the number of snapshots $n_s$ used to construct the compact model also affects its convergence to the full solution. Convergence is achieved more efficiently using a high number of samples in the snapshots matrix $S$ than using a small number of samples. The plots in Fig. 6 illustrate the convergence of the reduced solution using 185, 100, 70 and 40 snapshots while fixing the order of $k$ = 7 and interpolation order $m_1=m_2$ = 5.

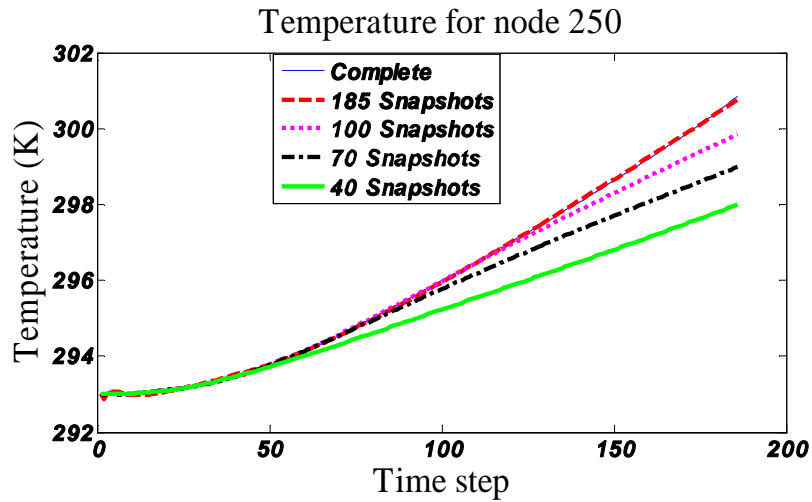

Figure 6. Convergence of the reduced solution by increasing the number of snapshots for order of basis *k=7* and interpolation points with *m₁=m₂* = 5.

*4.1 Validation of the reduced model using simulated irradiance data*
In this section, we investigate the sensitivity of the reduced model to the input irradiance signal. To this end, the reduced model is first constructed based on snapshots derived from experimental irradiance data, see Fig. 3. Then, simulated irradiance data is used to excite the model, and the corresponding response is compared to the full system response computed by direct time discretization. The irradiance data for this validation was carefully selected to have different environmental characteristics with the identification experimental data. On this basis, a day is chosen in autumn of November, 2011 with average air temperature of 22°C and average wind speed of 3 m/s. It is verified that the reduced model approximates the full system to a reasonable degree of accuracy also under this different excitation, as shown in Fig. 7.

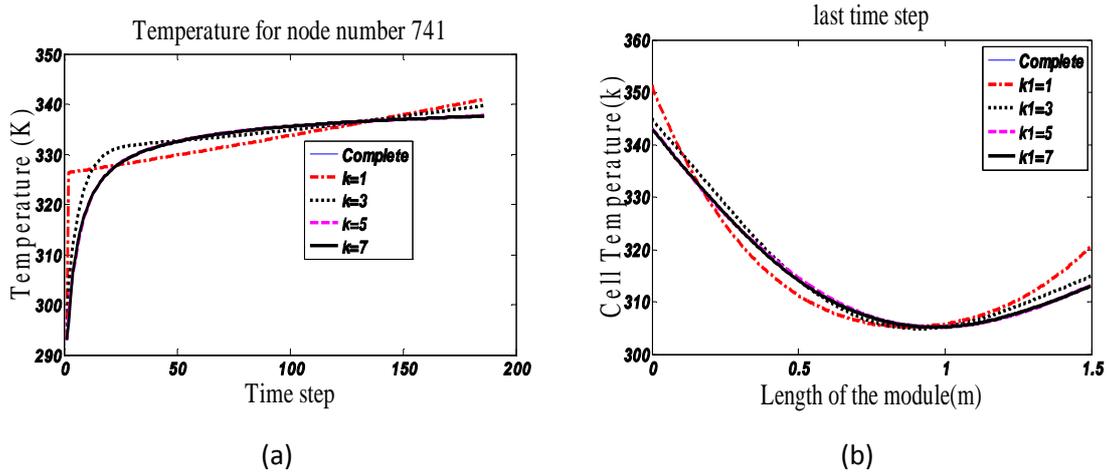

(a)              (b)

Figure 7. Convergence of reduced solution to the full solution with increasing order of *k* and fixed interpolation points $m_1=m_2 = 3$ for (a) a specific point in the silicon layer, and (b) for the entire silicon layer.

*4.2 Error analysis*

To demonstrate the efficiency and accuracy of the nonlinear order reduction, an error plot is deemed necessary to observe the convergence of the responses by varying the interpolation points $m_1$, $m_2$ and the dynamical order $k$. To do this, we compute a normalized error $\varepsilon(k, m_1, m_2)$ as

$$\varepsilon = \frac{\|T - \tilde{T}\|_2}{\|T\|_2} \tag{48}$$

where the norm is defined either in time domain by fixing the cell location, or in the space domain by fixing time step.

In order to observe the rate of convergence of the reduced system as we increase its dynamical order *k* with a fixed number of interpolation points $m_1=m_2=m$, we compute the error at specific time steps representative of the beginning, middle and end of simulation. A time variation error plot is also obtained for a selected element in the PV module for all the time steps. The results are shown in Fig. 8.

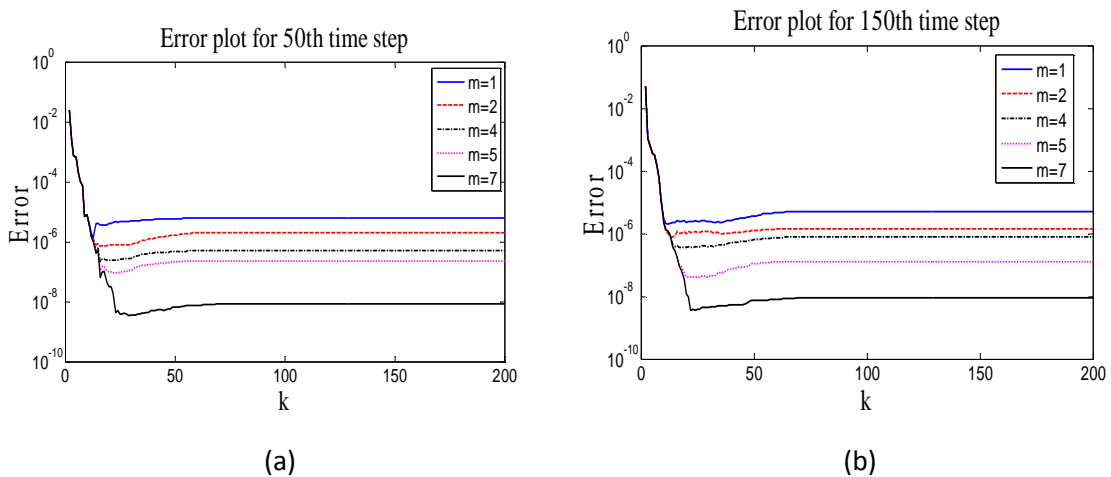

(a)              (b)

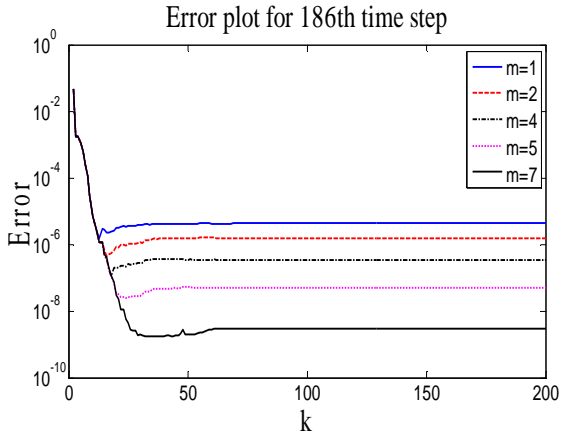
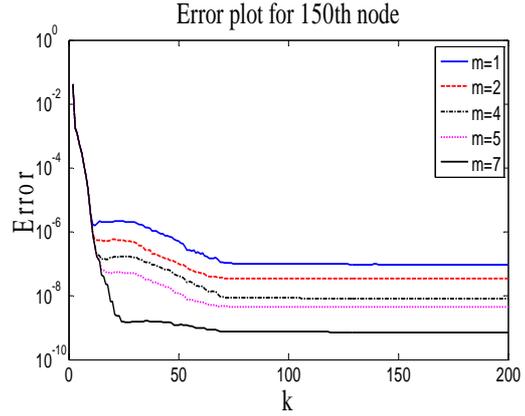

(c)                                                 (d)

Figure 8. Error plot for the topmost layer for fixed $m_1=m_2=m$ at (a) 50th time step (b) 150th time step (c) 186th time step, and for (d) node number 150 in the discretized module for all 186 time steps.

A clear lower error bound can be observed from the error plots, which is an indication that the reduced system response converges, as the order of the system is increased, only to the extent allowed by the representation of the nonlinear terms. It should be generally noted that the error obtained by using only one interpolation point (i.e. an equivalent of a linear system) is in the range of $10^{-5}$ to $10^{-6}$, an indication that the nonlinearity in the system is not at all strong. Increasing the number of interpolation points further shifts the error bound from $10^{-5}$ to less than $10^{-8}$ which confirms the excellent suitability of the DEIM algorithm for this thermal modelling task.

In order to verify the improved efficiency that can be achieved by using different interpolation points for the two nonlinear terms, as against using the same interpolation order for the two nonlinear terms, we perform various experiments by independently varying $m_1$ and $m_2$. The results are shown in Fig. 9.

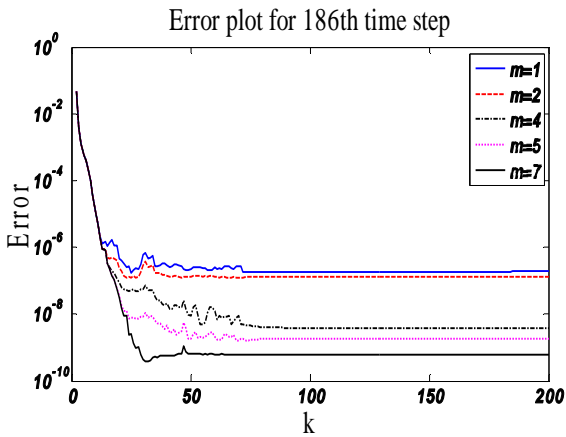
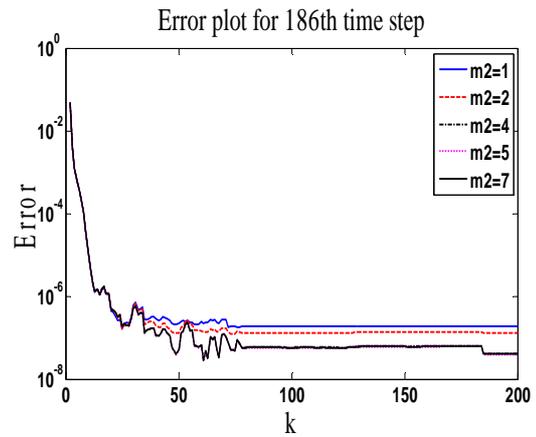

(a)                                                 (b)

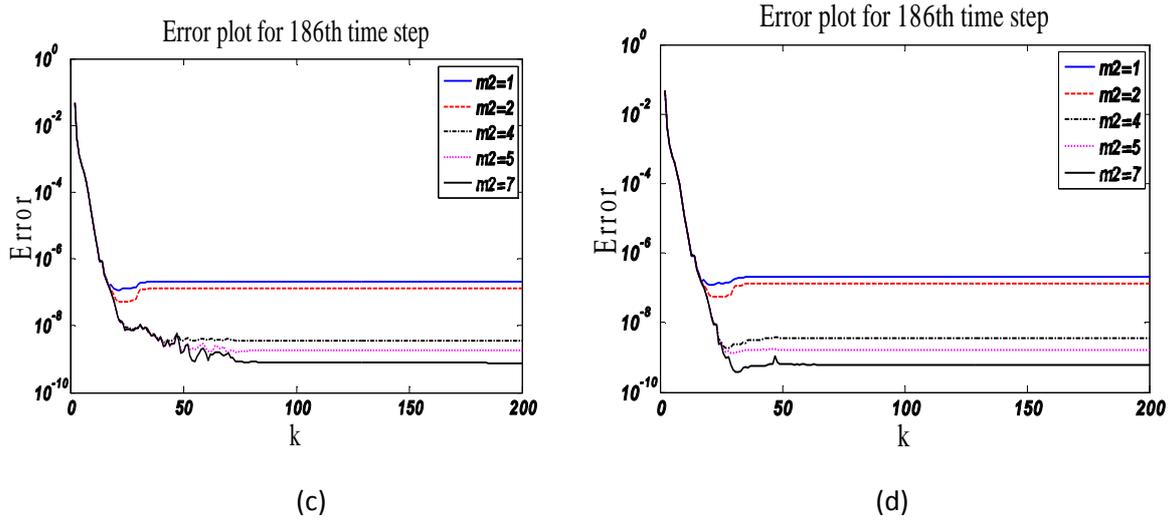

(c)                                                (d)

Figure 9. Error plot for the third layer at the 186th time step for (a) $m_1=m_2=m$ (b) $m_1=1$ (c) $m_1=5$ (d) $m_1=7$

The plots shown in Fig. 9 confirm that by independently varying $m_1$ and $m_2$, we can achieve better control of the accuracy of the reduced model. In Fig. 9a where the same interpolation order is used for the two nonlinear terms, the reduced model becomes efficient as order $k$ is increased above 50 when the lower bound error becomes more stable. By varying $m_1$ and $m_2$ independently, the stability of the lower error bound is attained with a smaller order $k$. It can be clearly observed in Fig. 9b-9d that there is a reduction in the lower bound error from less than $10^{-6}$ to less than $10^{-8}$ as $m_1$ is increased from 1 to 7 while $m_2$ is varied for each fixed $m_1$. Furthermore with independent variation of $m_1$ and $m_2$, a lower order $k$ (< 50) of the linear subspace projection is required to attain a stable error bound. This observation proves that, a better approximation of the full system can be achieved by independent variation of the interpolation of the nonlinear terms.

*4.3 Runtime*

Finally, we emphasize the advantages of proposed MOR technique by reporting the runtime required for the various simulations on the same commodity laptop. A transient analysis of the full system requires 61.40 seconds. Based on available snapshots, the construction of the reduced model via the proposed POD/DEIM requires as few as 0.86 seconds, whereas the transient simulation of the reduced model (k=7 and $m_1=m_2=3$) takes only 1.40 seconds. Excluding model setup and construction, the overall speedup is almost 44X. Considering that the proposed model is a quite simplified and 2D structure, more dramatic speedup is expected when applying this process to a full 3D geometry.

**5. Conclusion**

The partial differential equation governing the heat conduction within a 2D photovoltaic system has been derived and a numerical solution based on a finite difference scheme has been proposed. In order to limit the computational burden, we have identified the DEIM as a suitable technique to reduce the system through a combination of linear subspace projection and interpolation of nonlinear terms. In the numerical examples, we show that the heat conduction of a PV system discretized into 2166 nodes along the module span is successfully reduced to a compact model with dynamical order $k$=7, based on interpolation with only $m_1$=$m_2$=3 points. For the specific system under investigation, we see that further increment in $k$ does not result into a significant accuracy, and that the efficiency in the numerical solution of the reduced model, which depends on the number of interpolation points, can be fine- tuned by carefully selecting a different interpolation order for individual nonlinear terms. From the validation results, it is also concluded that the reduced solution is not very sensitive to the input function as it approximates well the simulated irradiance data much the same way as the experimental irradiance data used in the construction of the compact model. As a future development of the present research, the generalization of the present formulation for coupled thermo-mechanical problems is envisaged. In this case, the presence of cohesive crack bi-material interfaces between the layers, to model temperature-induced debonding of the back-sheet, should be exploited in order to assess the durability of PV modules. Also, extension to a full 3D geometry is under way.

**Acknowledgements**

MP would like to thank the European Research Council for the ERC Starting Grant "Multi-field and multi-scale Computational Approach to Design and Durability of PhotoVoltaic Modules" (ERC Grant Agreement n. 306622 - CA2PVM) granted under the European Union's Seventh Framework Programme (FP/2007-2013). OSO would like to acknowledge the support of the Italian Ministry of Education, University and Research to the project FIRB 2010 Future in Research "Structural mechanics models for renewable energy applications" (RBFR107AKG).

**REFERENCES**

[1] M. Paggi, I. Berardone, A. Infuso, M. Corrado, Fatigue degradation and electric recovery in Silicon solar cells embedded in photovoltaic modules. *Scientific Reports* 4 (2014) 4506 doi:10.1038/srep04506

[2] Eitner, U., Köntges, M., and Brendel, R. Use of digital image correlation technique to determine thermomechanical deformations in photovoltaic laminates: measurements and accuracy. Solar Energy Mater. Solar Cells, 94 (2010) 1346–1351.

[3] M. Köntges, I. Kunze, S. Kajari-Schröder, X. Breitenmoser, B. Bjrneklett, The risk of power loss in crystalline silicon based photovoltaic modules due to microcracks. Sol. Energy Mater. Sol. Cells, 95 (2011) 1131–1137.

[4] A. Sapora, M. Paggi, A coupled cohesive zone model for transient analysis of thermoelastic interface debonding, Computational Mechanics, 53 (2014) 845-857.

[5] M. Paggi, M. Corrado, M.A. Rodriguez. A multi-physics and multi-scale numerical approach to microcracking and power-loss in photovoltaic modules. Compos Struct 95 (2013) 630-638.

[6] Y. Leea, A. Andrew, B. Taya, Finite Element Thermal Analysis of a Solar Photovoltaic Module, Energy Procedia 15 (2012) 413 – 420.

[7] M. Paggi, A. Sapora, Numerical Modelling of Microcracking in PV Modules induced by Thermo-mechanical loads, Energy Procedia 38 ( 2013 ) 506–515.

[8] A. C. Antoulas, Approximation of Large-Scale Dynamical Systems. Society for Industrial and Applied Mathematics, Philadelphia, PA, USA, 2005.

[9] Peter Benner, Michael Hinze, and E Jan W Ter Maten. Model reduction for circuit simulation, Lecture Notes in Electrical Engineering, vol. 74, Springer, 2010

[10] S.X.D Tan and L. He. Advanced Model Order Reduction Techniques in VLSI Design. Cambridge University Press, New York, 2007

[11] W.H.A. Schilders, H.A. Van Der Vorst, and J. Rommes. Model order reduction: theory, research aspects and applications. Springer Verlag, 2008

[12] A. Odabasioglu, M. Celik, and L. T. Pileggi, "PRIMA: passive reduced-order interconnect macromodeling algorithm," IEEE Trans. on Computer-Aided Design of Integrated Circuits and Systems, vol. 17, no. 8, pp. 645–654, August 1998.

[13] W.E. Arnoldi. The principle of minimized iterations in the solution of the matrix eigenvalue problem. Quart. Appl. Math, 9(1):17–29, 1951.

[14] C. Lanczos. An iteration method for the solution of the eigenvalue problem of linear differential and integral operators1. Journal of Research of the National Bureau of Standards, 45(4), 1950.

[15] B. Moore. Principal component analysis in linear systems: Controllability, observability, and model reduction. Automatic Control, IEEE Transac- tions on, 26(1):17 – 32, feb 1981

[16] K. Glover. All optimal hankel-norm approximations of linear multivari- able systems and their l—inf error bounds. International Journal of Control, 39(6):1115–1193, 1984

[17] K. Willcox and J. Peraire. "Balanced model reduction via the proper orthogonal decomposition." AIAA journal, vol. 40, no. 11 (2002): 2323-2330.

[18] B. Bond, Z. Mahmood, Y. Li, R. Sredojevic, A. Megretski, V. Stojanovic, Y. Avniel, and L. Daniel, "Compact modeling of nonlinear analog circuits using system identification via semidefinite programming and incremental stability certification," IEEE Trans. on CAD of Integrated Circuits and Systems, vol. 29, no. 8, pp. 1149–1162, 2010.

[19] J. R. Phillips, "Projection-based approaches for model reduction of weakly nonlinear, time-varying systems," IEEE Trans. Comput.-Aided Design Integr. Circuits Syst., vol. 22, no. 2, pp. 171–187, Feb. 2003

[20] M. Rewienski and J. White, "A trajectory piecewise-linear approach to model order reduction and fast simulation of nonlinear circuits and micro-machined devices," IEEE Trans. Comput.-Aided Design Integr. Circuits Syst., vol. 22, no. 2, pp. 155–170, Feb. 2003

[21] Astolfi, A., "Model Reduction by Moment Matching for Linear and Nonlinear Systems," Automatic Control, IEEE Transactions on, vol.55, no.10, pp.2321,2336, Oct. 2010

[22] S. Chaturantabut, D. Sorensen, Nonlinear Model Reduction via Discrete Empirical Interpolation, SIAM J. SCI. COMPUT, Vol. 32, No. 5 (2010), 2737–2764.

[23] A. Jones C. Underwood, A Thermal Model for Photovoltaic Systems, Solar Energy Vol. 70, No. 4 (2001) 349–359.

[24] S. Armstrong, W. Hurley, A Thermal Model for Photovoltaic Panels under Varying Atmospheric Conditions, Applied Thermal Engineering 30 (2010) 1488-1495.

[25] T. Blomberg, Heat Conduction in Two and Three Dimensions: Computer Modelling Of Building Physics Applications, Report TVBH-1008, Department of Building Physics, Lund University, Sweden (1996), 5-13.